\documentclass[useAMS,usenatbib]{mnras}
\usepackage{graphicx}
\usepackage{txfonts}
\usepackage[dvipsnames]{color}
\usepackage{ulem}
\usepackage{ragged2e}
\def\aap{A\&A}
\def\apj{ApJ}
\def\apjl{ApJL}
\def\apjs{ApJS}
\def\mnras{MNRAS}

\def\nat{Nature}

\usepackage{soul}
\usepackage[dvipsnames]{xcolor}

\title[RRMSs in photon starved regime]{Monte-Carlo simulations of relativistic radiation mediated shocks: II. photon starved regime}
\author[Ito et al.]{{Hirotaka Ito$^{1,2}$, Amir Levinson$^{3}$, and Shigehiro Nagataki$^{1,2}$} \\
  $^{1}$Astrophysical Big Bang Laboratory, RIKEN, Saitama 351-0198, Japan; hirotaka.ito@riken.jp\\
  $^{2}$Interdisciplinary Theoretical \& Mathematical Science Program (iTHEMS), RIKEN, Saitama 351-0198, Japan\\
$^{3}$School of Physics \& Astronomy, Tel Aviv University, Tel Aviv 69978,
  Israel; levinson@wise.tau.ac.il\\
  }

\begin{document}
\date{\today}
\pagerange{000--000} \pubyear{0000}
\maketitle
\label{firstpage}
\begin{abstract}
Radiation mediated shocks (RMS) play a key role in shaping the early emission observed in many transients.
In most cases, e.g., shock breakout in supernovae, llGRBs and neutron star mergers, the upstream plasma is devoid of radiation, 
and the photons that ultimately reach the observer are generated predominantly inside and downstream of the shock.  
Predicting the observed spectrum requires detailed calculations of the shock structure and thermodynamic state that account properly for the shock microphysics. 
We present results of self-consistent Monte-Carlo simulations of photon-starved RMS, that yield the shock structure and emission for a broad range of shock velocities, 
  from  sub-relativistic ($\beta_{sh} = 0.1$) to highly relativistic ($\Gamma_{sh} = 20$).  
  Our simulations confirm that in relativistic RMS the immediate downstream temperature is regulated by exponential pair creation,
  ranging from $50$ keV at $\beta_{sh}=0.5$ to $200$ keV at $\Gamma_{sh}=20$.
 At lower velocities the temperature becomes sensitive to the shock velocity, with $kT\sim 0.5$ keV at $\beta_{sh}=0.1$.
 We also confirm that in relativistic shocks the opacity is completely dominated by newly created pairs, which has important implications for the breakout physics.  
 We find the transition to pair dominance to occur at $\beta_{sh}=0.5$ roughly.
  In all cases examined, the spectrum below the $\nu F_\nu$ peak has been found to be substantially softer than the Planck distribution. This has important implications
for the optical emission in fast and relativistic breakouts, and their detection.   The applications to GRB 060218 and GRB 170817A are discussed.
\end{abstract}


\section{Introduction}
\label{Intro}
The early emission observed in various types of cosmic explosions is released during the breakout of a radiation mediated
shock (RMS) from the envelope enshrouding the blast center \citep[for a recent review, see][]{LN19}.
In such shocks the dissipation mechanism involves emission and scattering of radiation and, under certain
conditions, pair production.   The properties  of the breakout signal and the subsequent cooling envelope
emission depend on the shock velocity and structure, as well as on the upstream conditions. 
The shock velocity during the breakout phase can range from subrelativistic in regular supernovae, through mildly relativistic in, e.g.,
long gamma-ray bursts (LGRBs) and neutron star (NS) mergers, to highly relativistic in energetic aspherical explosions of compact progenitors \citep{NS12}.

Two disparate regimes have been identified  \citep{LN19}; photon rich shocks in which advection of upstream radiation determines the downstream state,
and photon starved shocks in which photons are generated predominantly inside and just downstream of the shock transition layer.  The former case
is anticipated in sub-photospheric shocks in GRBs \citep{LB08,LE12,B17,LBV18,ILS18}, whereas the latter in most other systems 
\citep{W76,KBW10,BKSW10,NS10,NS12,GNL18,IL19}.
In a previous paper \citep[][hereafter paper I]{ILS18} we presented a comprehensive analysis of photon rich RMS and its application to LGRBs, using a
Monte-Carlo method that solves the shock structure coupled to the transfer of radiation through the shock in a self-consistent manner.
We have shown that photon advection dominates over photon generation when the photon-to-baryon ratio far upstream exceeds the e-p mass
ratio times the shock Lorentz factor, and that this ratio determines the immediate downstream temperature.  We have also shown that the 
observed spectrum is expected to be broad owing to bulk comptonization.  Similar results were obtained by \cite{B17} and \cite{LBV18} using
a different method, verifying the validity of the two techniques. 

In this paper we use a modified version of our Monte-Carlo code (see Appendix \ref{sec:appA} for details), that includes photon generation and absorption in addition to scattering and pair creation, to compute the structure and spectrum of photon starved shocks for a broad range of shock velocities, from $\beta_u=0.1$ up to $\Gamma_u=20$,
where $\beta_u$ is the velocity of the upstream plasma, as measured in the shock frame, and $\Gamma_u$ is the corresponding Lorentz factor. 

In general, photon starved RMS exhibit three different velocity domains with vastly different behaviours  \citep{KBW10,LN19}:  
(i) Slow shocks ($\beta_u<0.05$) in which the radiation is in full thermodynamic equilibrium already inside the shock transition
layer, and the temperature depends weakly on shock velocity and plasma density.  In such shocks the emitted spectrum is a black body spectrum.  (ii) Fast
Newtonian shocks ($0.05 \lesssim \beta_u \lesssim 0.5$), in which the radiation is out of thermodynamic equilibrium, and the temperature is determined by the 
photon generation rate in the immediate downstream, and depends very sensitively on shock velocity. 
(iii) Relativistic shocks 
in which pair creation and Klein-Nishina effects play a dominant role.   Relativistic RMS have been thoroughly analyzed in the highly relativistic regime 
($\Gamma_u \ge6$) by \cite{BKSW10}, using a kinetic approach to solve the radiative transfer equation inside the shock. 
Our analysis, that exploits a different technique, confirms those previous results, and is extended to the fast Newtonian and mildly relativistic 
regimes that apply to the majority of strong stellar explosions, allowing, for the first time, to perform self-consistent calculations of the
RMS structure and inherent spectrum in these cases, which is the prime goal of this study.
The shock structure computed in our simulations is found to be in good agreement with semi-analytic solutions obtained both in the  
Newtonian limit \citep{BP81b,IL19} and in the highly relativistic limit (\citealt{NS12,GNL18} hereafter GNL18).
More importantly, we find that in fast Newtonian and relativistic RMS the spectrum during shock breakout (and conceivably the early cooling 
emission) deviates considerably from a black body, which has important  implications for the detection of these sources.

 This paper is organized as follows.   In Section \ref{Numerical} we describe the numerical method and the setup of  our simulations.
We present the main results in Section \ref{result}. In Section \ref{application} we discuss the applications to shock breakouts, with particular attention 
to GRB 060218 and GRB 170817A. We conclude in Section \ref{conclusion}.
Throughout the paper, the subscript $u$ and $d$  refer to the physical quantities at the far upstream and 
far downstream  regions of the shock, respectively.

\section{Numerical Setup}
\label{Numerical}

The Monte-Carlo RMS code which we have developed enables us to compute the steady-state profile of RMS
 for a  range of shock velocities that encompasses the sub-relativistic and ultra-relativistic regimes. 
The code iteratively seeks a self-consistent flow profile that satisfies energy-momentum conservation at all grid points.  More precisely, 
under the assumption that protons and pairs each have a local Maxwellian distribution with the same temperature, we compute the transfer of 
radiation through the shock for the given plasma profile using Monte-Carlo techniques. 
Based on the departure from the energy-momentum conservation 
evaluated in each cycle, the shock profile is modified iteratively until convergence is reached to the desired accuracy.
The details are described in Paper I for photon rich RMS.
The main modification in the present analysis is the inclusion of
free-free emission/absorption processes which are essential in photon starved shocks.
A summary of the updates is outlined in Appendix \ref{sec:appA}.

In the current study, the input parameters are the 4-velocity of the upstream flow, $\Gamma_{\rm u}\beta_{\rm u}$, the proper baryon density at the far upstream region  $n_{\rm u}$,\footnote{Unlike in the photon rich regime, the photon-starved RMS solution (i.e., shock profile expressed as a function of optical depth) has explicit dependence on the absolute value of $n_{\rm u}$ due to the inclusion of the free-free absorption process.} and the composition, which we take to be purely hydrogen.
As for the baryon density, we invoke a fixed value of $n_{\rm u} = 10^{15}~{\rm cm}^{-3}$ which is identical to that adopted in the calculations of \cite{BKSW10}. It is worth noting that the
solutions are not sensitive to the choice of  $n_{\rm u}$;
in the fully and mildly relativistic regimes the immediate downstream temperature is always around $\sim 100-200~{\rm keV}$, owing to 
the pair creation thermostat \citep[see ][and next section]{BKSW10}. Even in the sub-relativistic regime the 
dependence of the temperature on $n_{\rm u}$ is only mild \citep[but might be important from an observational perspective, see][for discussion]{LN19}.
The composition can alter the downstream temperature if heavy and in particular in case of r-process material \citep[see Fig. 15 in][]{LN19}; we leave the
investigation of composition effects, that are mainly relevant to NS mergers, for a future work. 

As stated in the introduction, we are interested in exploring the regimes of fast Newtonian and relativistic shocks ($\beta_{\rm u} >0.05$).
To that end we consider 6 models with different values of the upstream 4-velocity, which translate to $\beta_{\rm u} = 0.1$, $0.5$, and $\Gamma_{\rm u} = 2$, $6$,  $10$ and $20$.
The Lorentz factors in the highly relativistic regime ($\Gamma_{\rm u} = 6, 10, 20$) were
chosen to enable a direct comparison with the results of \citet{BKSW10}.
As for the remaining cases, to our knowledge this is the first time ever that ab-initio calculations of RMS in those regimes have been performed. 

As described in Appendix \ref{sec:app_boundary}, we inject a minuet amount of thermal photons at the upstream boundary  merely for numerical convenience. 
This has no practical effect on the solutions, since the new photons generated by free-free emission dominate the radiation field already in
the far upstream.   To confirm this, we checked that the results converge by changing the density and energy of the injected photon population.

\section{Results}
\label{result}

\subsection{Overall structure}
\label{Overall}

\begin{figure}
\begin{center}
\includegraphics[width=8cm,keepaspectratio]{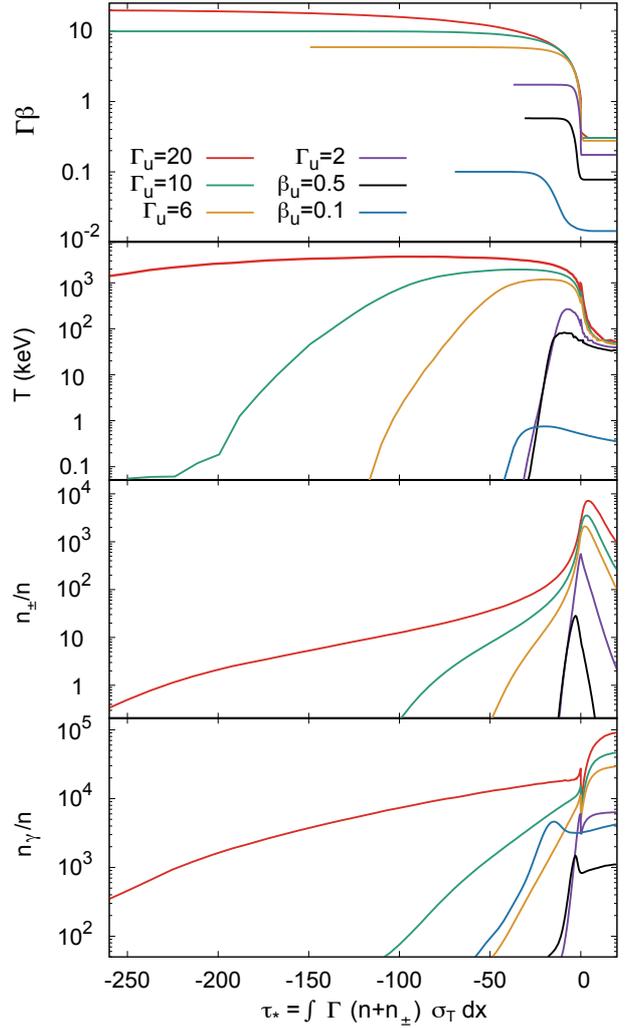} 
\end{center}
\caption{Overall shock structure for upstream velocities of $\beta_{\rm u} = 0.1$ ({\it blue}),  $\beta_{\rm u} = 0.5$ ({\it black}),  $\Gamma_{\rm u} = 2$ ({\it magenta}),  $\Gamma_{\rm u} = 6$ ({\it brown}),  $\Gamma_{\rm u} = 10$ ({\it green}) and   $\Gamma_{\rm u} = 20$ ({\it red}).
 In each panel, from top to bottom, we display the  4-velocity $\Gamma \beta$, the plasma
 temperature $T$,
  the pair -to- baryon density ratio $n_{\pm}/n$ and  the photon -to- baryon ratio $n_{\gamma}/n$, as a function of optical depth $\tau_{*}$.
 The location of $\tau_{*} = 0$ are taken at the position of the subshock ($\Gamma_{\rm u} \geq 2$) or the position where the velocity satisties $\beta = 1.05 \beta_{\rm d}$ when subshock is absent ($\beta_{\rm u} \leq 0.5$).}
\label{ALL}
\end{figure}

Fig.  \ref{ALL} shows the profiles of the 4-velocity, temperature, pair-to-baryon ratio and photon-to-baryon ratio,  plotted as functions of the angle averaged, pair loaded
Thomson optical depth, $\tau_{*} = \int \Gamma (n + n_{\pm}) \sigma_T dx$, where $n_\pm$ is the pair density and $\sigma_T$ is the 
Thomson cross section.  A zoom in of the immediate downstream region is shown in 
Fig. \ref{SUB} for the relativistic shocks. It indicates formation of subshocks, as also found in \cite{BKSW10}.  While non negligible, they do not 
alter significantly the overall shock structure.  We shall discuss them  in greater detail below.   Such subshocks 
are absent in the fast Newtonian regime ($\Gamma_{\rm u} \beta_{\rm u}< 1$).

Fig. \ref{ALL}  confirms that the  transition from the fast Newtonian regime to the relativistic regime occurs at $\beta_{\rm u}\simeq0.5$.
The temperature just behind the shock changes from about $0.5$ keV at $\beta_{\rm u}=0.1$ to about $50$ keV at $\beta_{\rm u}=0.5$, 
confirming the sensitive dependence found earlier analytically \citep{W76,KBW10}.   At $\beta_{\rm u}=0.1$ the simulation result is in excellent agreement 
with the analytic estimates, as can be seen by comparing the value found in our simulation with Fig 5 in \cite{LN19}.  At $\beta_{\rm u}>0.5$ there is only a very weak dependence 
of the temperature on $\Gamma_{\rm u}\beta_{\rm u}$, owing to the exponential pair creation thermostat mentioned above.   The rapid increase of the pair
content with increasing $\Gamma_{\rm u}\beta_{\rm u}$ is clearly seen in the bottom panel of Fig.  \ref{ALL}
 for $\beta_{\rm u}>0.5$.  
At lower velocities the $n_\pm/n$ ratio is found to be practically zero.
    The photon generation along the flow
    results in a significant  increase in the photon number towards downstream.
    As seen in the figure, emergence of copious pairs  for $\beta_{\rm u} \geq 0.5$  
    largely enhances the photon production. It is noted that the discontinuous change of photon to baryon ratio at $\tau_* = 0$ for $\Gamma_{\rm u} \geq 2$ 
    is due to the instantaneous change in the inertial frame as well as the comoving baryon density across the subshock.

As also seen,  the shock width increases with decreasing velocity in the Newtonian regime, whereas it increases with increasing Lorentz 
factor in the relativistic regime.  As will be discussed in more detail in Section \ref{WIDTH} below, the reason for this apparently peculiar behaviour is that 
in the Newtonian regime the width is set by the diffusion length of photons, whereas in the relativistic regime it is largely
affected by Klien-Nishina suppression.   
This suppression allows photons that propagate from the downstream to the upstream 
to penetrate to a much larger distances ahead of the shock and, 
as a result, a substantial increase in the temperature and pair density begins well before the flow decelerates, at increasingly larger distance for larger $\Gamma_{\rm u}$.
The maximal value of the temperature is attained at the upstream, while the pairs density reaches its maximum in the near downstream (Fig. \ref{SUB}).
A quantitative scaling of the physical shock width across the entire velocity range 
is derived in Section \ref{WIDTH} below. 

\subsubsection{Subshocks}
\label{SUBSH}

\begin{figure}
\begin{center}
\includegraphics[width=8cm,keepaspectratio]{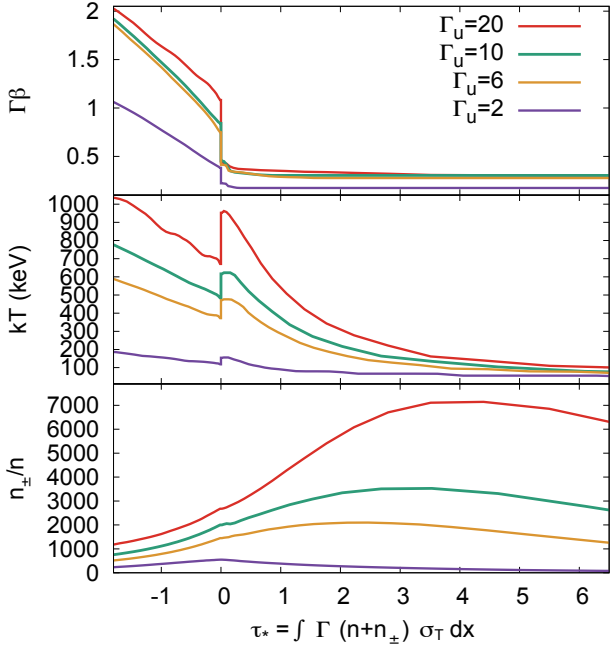}
\end{center}
\caption{Enlarged view of the 4-velocity ({\it top}), temperature ({\it middle}) and pair -to- baryon ratio ({\it bottom}) near the subshock
  for upstream Lorentz factors of
  $\Gamma_{\rm u} = 2$ ({\it magenta}),  $\Gamma_{\rm u} = 6$ ({\it brown}),  $\Gamma_{\rm u} = 10$ ({\it green}) and   $\Gamma_{\rm u} = 20$ ({\it red}).
 }
\label{SUB}
\end{figure}

As mentioned above, one of the characteristic features which is only seen in relativistic RMS
 is formation of a subshock.
 Our simulations indicate that, while  the photon-plasma interaction leads to a smooth transition at $\beta_{\rm u} \lesssim 0.5$, 
 subshocks\footnote{The subshock is presumably mediated by collective plasma processes on kinetic scales that are 
  much shorter than the mean free path of photons. In our analysis it is treated as a discontinuity in the flow parameters 
across which the Rankine-Hugoniot jump conditions are satisfied.}  inevitably form when $\Gamma_{\rm u} \gtrsim 2$.  A similar 
phenomena was found also in photon rich shocks, although the subshocks there are much weaker (paper I).
A close-up view of the subshock region is shown in Fig. \ref{SUB}. It is overall consistent with the substructures seen 
 in the simulations of \citet[][]{BKSW10}, however, the strength of the subshocks in our simulations are larger  than those reported in  \citet[][]{BKSW10}.
 The velocity jump across the subshock they find is roughly $\delta (\Gamma \beta) \sim 0.1$ for all the cases they explored ($\Gamma_{\rm u} = 6, 10, 20$ and $30$),
 implying negligible energy dissipation,
 whereas we find velocity jumps of 
  $\delta (\Gamma \beta) \sim 0.16$, $0.33$, $0.38$ and $0.66$ 
 for $\Gamma_{\rm u} = 2$, $6$, $10$ and $20$, respectively,
 with subshock dissipation of 
a few  percents
of the total shock energy\footnote{The total kinetic energy flux that is dissipated in the shock is given as
 $J m_p c^3 (\Gamma_{\rm u} - \Gamma_{\rm d})$, where $J = \Gamma_{\rm u} n_{\rm u} \beta_{\rm u}$ is the baryon number flux.
    Taking into account the presence of pairs, the energy dissipated in the subshock is given by
    $J [m_p + (n_{\pm}/n)_{\rm sub}m_e]c^3 (\Gamma_{\rm u, sub} - \Gamma_{\rm d, sub})$, where $(n_{\pm}/n)_{\rm sub}$ is the pair to baryon ratio at the subshock and   $\Gamma_{\rm u, sub}$ and $\Gamma_{\rm d, sub}$ are
    the Lorentz factors of the flow at immediate upstream and downstream of the subshock, respectively.  Hence, the fraction of energy dissipated in the subshock can be estimated as $[1+(n_{\pm}/n)_{\rm sub}m_e/m_p](\Gamma_{\rm u, sub} - \Gamma_{\rm d, sub})/(\Gamma_{\rm u} - \Gamma_{\rm d})$.
}
%
The origin of this discrepancy may be traced to the approximation imposed in their analysis.
A careful scrutiny of their analysis can be found in Appendix \ref{DIFFERENCE}.
Apart from these details, our simulations show a broad agreement with \citet{BKSW10}, as will be discussed further below.

\begin{figure*}
\begin{center}
\includegraphics[width=15cm,keepaspectratio]{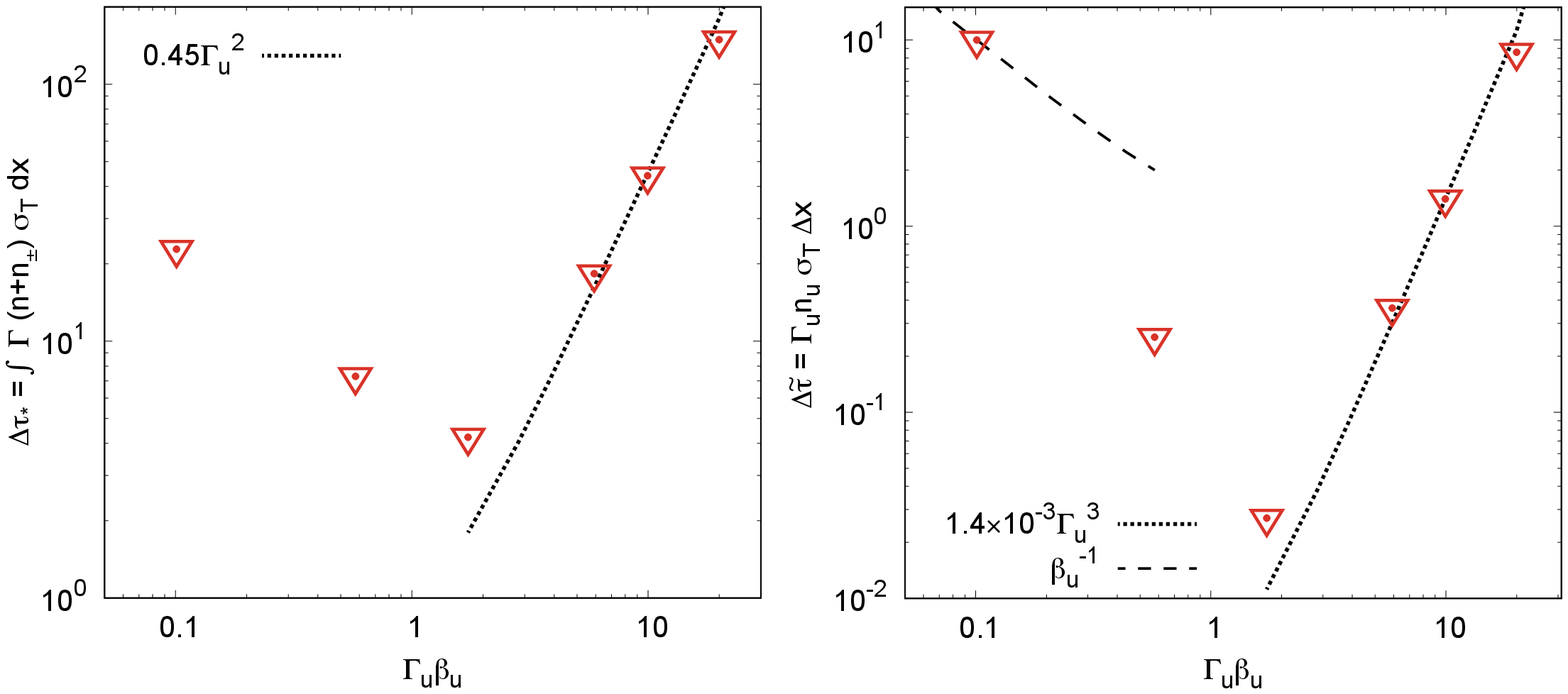}
\end{center}
\caption{Pair loaded Thomson optical depth (left) and dimensionless width (right) of the shock transition layer. 
  Here the shock width $\Delta x$, as measured in the shock frame, is defined as the distance from the location where $(\Gamma \beta)/(\Gamma_{\rm u} \beta_{\rm u}) = 0.9$ to
  the subshock, or to the downstream point where   $(\Gamma \beta)/(\Gamma_{\rm d} \beta_{\rm d}) = 1.1$ when a subshock is absent. 
The red triangles show the results obtained from the simulations.  The dotted and dashed lines delineate the scaling anticipated in the highly relativistic 
( $\Delta \tau_*\propto\Gamma_{\rm u}^2$ and $\Delta\tilde{\tau}\propto \Gamma_{\rm u}^3$) and subrelativistic ($\Delta\tilde{\tau}\propto \beta_{\rm u}^{-1}$) regimes, respectively.
}
\label{TAUBETA}
\end{figure*}

\subsubsection{Scaling of the shock width}
\label{WIDTH}
 
 As stated above, the width of the shock transition layer (i.e., the deceleration zone)  is a non-monotonic function of the shock 4-velocity.
In the non-relativistic regime ($\beta_{\rm u} \ll 1$) the transport of radiation across the shock is diffusive, and the transition occurs rather 
gradually over one diffusion length roughly \citep{W76,BP81a,BP81b,KBW10}.  By equating the diffusion time across the shock, $t_D=\Delta \tau_* L_s/c$,
here $L_s$ is the shock width and $\Delta\tau_*=\int_{-L_s}^0\sigma_Tndx$ is the optical thickness of the shock,
with the flow time, $t_f=\int_{-L_s}^0dx/c\beta  =\Delta \tau_*/(\sigma_Tn_{\rm u}\beta_{\rm u})$ (recalling that $n\beta =n_{\rm u}\beta_{\rm u}$),
one obtains: 
\begin{equation}
\sigma_T n_{\rm u} L_s \approx 1/\beta_{\rm u}.
\label{eq:TAUNR}
\end{equation}
The right panel in Fig.  \ref{TAUBETA} shows good agreement between this naive estimate and the simulation result for $\beta_{\rm u}=0.1$ (see also Appendix \ref{NR}).

In difference, in the relativistic regime ($\beta_{\rm u} \sim 1$)  counterstreaming photons are mostly scattered back in a single scattering.\footnote{There is also contribution from pair production, but the opacity is smaller than that for Compton scattering.}
Nonetheless, the pair loaded Thomson optical depth is significantly larger than unity, and increases with increasing $\Gamma_{\rm u}$, by virtue of
Klein-Nishina effects.
In fact, the change of the shock width with $\Gamma_{\rm u}$ is nonlinear, since the temperature 
inside the shock is roughly proportional to the  local Lorentz factor (Fig. \ref{ALL}), implying that the mean photon energy seen in the rest frame 
of an electron (or positron) and, hence, the Klein-Nishina suppression, scale as $\Gamma^2$.\footnote{Note that since the temperature in the immediate 
downstream is fixed by pair creation, 
the mean energy of counterstreaming  photons, as measured in the shock frame, is roughly $m_ec^2$, independent of $\Gamma_u$, and
the local comoving energy is $\sim\Gamma m_ec^2$. About half of it is converted to internal energy (per lepton), hence the scaling.}
This heuristic result is in a good agreement with 
the simulations performed by \citet{BKSW10} and the analytic solution derived in 
\cite{NS12} and GNL18, who find the scaling  $\Delta \tau_{*}  \propto \Gamma_{\rm u}^2$ (up to a logarithmic factor).
These solutions also yield the scaling $\Delta \tilde{\tau}\propto \Gamma_{\rm u}^3$ for the pair unloaded 
depth of the shock transition layer, defined as $\Delta \tilde{\tau}=\int^0_{-\Delta x} \Gamma n \sigma_T dx=\sigma_T \Gamma_{\rm u} n_{\rm u} \Delta x$,
where $\Delta x$ is the length, as measured in the shock frame, over which the shock Lorentz factor changes substantially (see GNL18 for details).   
The optical thickness  $\Delta \tilde{\tau}$ corresponds to the minimum opacity needed to sustain the RMS;
once the total optical depth ahead of the shock becomes smaller than this value, viz., $\tau \lesssim \Delta \tilde{\tau}$, radiation starts leaking out of the shock
and the shock structure is significantly altered (GNL18).   This is the point where breakout commences. 

Fig. \ref{TAUBETA} shows the pair loaded Thomson depth of the shock transition layer $\Delta\tau_*$ (left panel), and the dimensionless shock width $\Gamma_{\rm u} n_{\rm u} \sigma_T \Delta x$ (right panel), 
measured in the simulations (the red triangles).   The latter equals the pair unloaded Thomson optical depth in the limit $\Gamma_{\rm u} \gg 1$.
The shock width $\Delta x$ is defined here as the backward distance (measured in the shock frame) from the subshock (or the point where $\Gamma\beta=1.1\Gamma_{\rm d}\beta_{\rm d}$
if there is no subshock), at which the 4-velocity reaches 90\% of its 
upstream value, that is,  $\Gamma(-\Delta x)\beta(- \Delta x)=0.9\Gamma_{\rm u}\beta_{\rm u}$.    As seen, while for $\beta_{\rm u}=0.1$ the shock thickness $\Delta \tilde{\tau}$ 
 agrees well with
Eq. (\ref{eq:TAUNR}), it is much narrower for $\beta_{\rm u}=0.5$.  The reason is that in this case  the opacity inside the shock is dominated by newly created pairs, as 
can be inferred by comparing the results for $\beta_{\rm u}=0.5$ in the left and right panels.   The simulation results also indicate that the scaling derived analytically in GNL18
and found numerically in \cite{BKSW10} holds
from $\Gamma_{\rm u}=6$ up to $\Gamma_{\rm u}=20$.
    At $\Gamma_{\rm u}=2$ we find somewhat  departure from this scaling.
    This is expected since the scaling is valid  in the relativitic limit.

The non-monotonic behaviour of the shock width implies that physical shock width ($\Delta x$ or equivalently $\Delta \tilde{\tau}$) has an absolute minimum.  From our simulations we estimate  that it occurs around $\Gamma_{\rm u}=2$ (see right panel in Fig. \ref{TAUBETA}).  

\begin{figure*}
\begin{center}
\includegraphics[width=17.5cm,keepaspectratio]{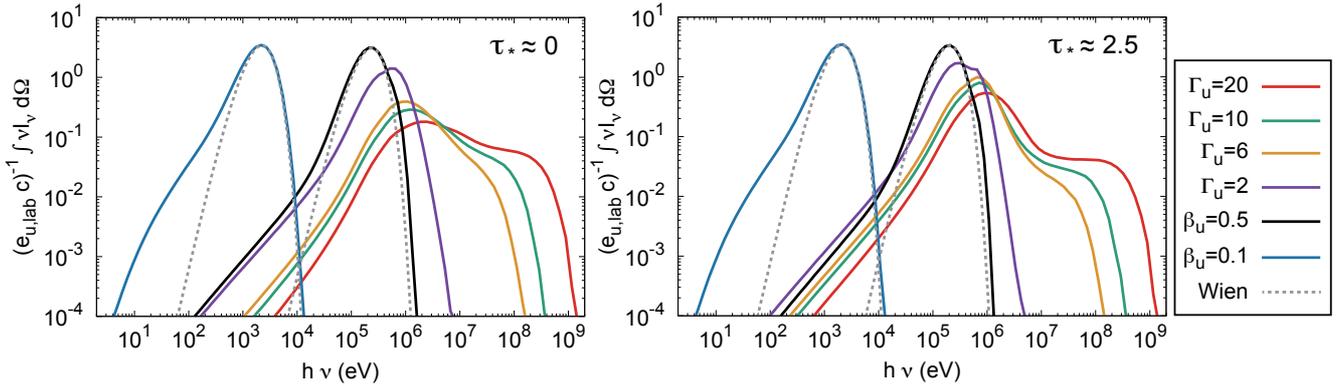}
\end{center}
\caption{Shock-frame, local, angle integrated SEDs, $c^{-1}\int \nu I_\nu d\Omega$, normalized by the total kinetic energy density of the far upstream flow, $e_{\rm u} = \Gamma_{\rm u} (\Gamma_{\rm u} - 1) n_{\rm u} m_p c^2$.
  The left and right panels correspond to the downstream locations $\tau_{*}\approx 0$ and  $\tau_{*}\approx 2.5$, respectively.  
  The blue, black, magenta, brown, green and red lines show the results for
  $\beta_{\rm u} = 0.1$,  $0.5$, and  $\Gamma_{\rm u} = 2$,  $6$,  $10$ and   $20$, respectively.  The dotted lines delineate the Wien spectra  for 
  the $\beta_{\rm u}=0.1$ and $\beta_{\rm u}=0.5$ cases.}
\label{SPDS}
\end{figure*}

\subsection{Spectrum}

\begin{figure}
\begin{center}
\includegraphics[width=8.5cm,keepaspectratio]{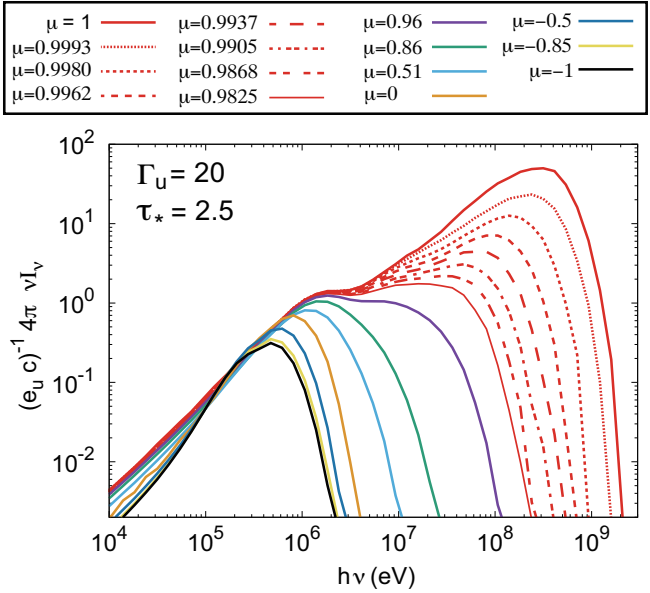}
\end{center}
\caption{Shock-frame, local SEDs ($c^{-1} 4\pi \nu I_\nu$)  at the downstream location $\tau_{*}=2.5$, for $\Gamma_{\rm u} = 20$.
The normalization is the same as in Fig. \ref{SPDS}.
  Each line shows the case for a given cosine of the angle between the direction of photon propagation and  the flow velocity, $\mu=\cos\theta$.
  The sequence of red lines demonstrates the strong angle dependence of the beamed high energy component  within a narrow range $\theta \lesssim 4 / \Gamma_{\rm u} \sim 11^{\circ}$.
  The magenta ({\it {$\theta=16^{\circ}$}}), green ({\it {$\theta=30^{\circ}$}}), cyan ({\it {$\theta=60^{\circ}$}}), brown ({\it {$\theta=91^{\circ}$}}), blue ({\it {$\theta=120^{\circ}$}}), yellow ({\it {$\theta=149^{\circ}$}}) and black lines ({\it {$\theta=180^{\circ}$}}) show the dependence well outside the beaming cone.  }
\label{SPDS_diff}
\end{figure}

\begin{figure*}
\begin{center}
\includegraphics[width=16cm,keepaspectratio]{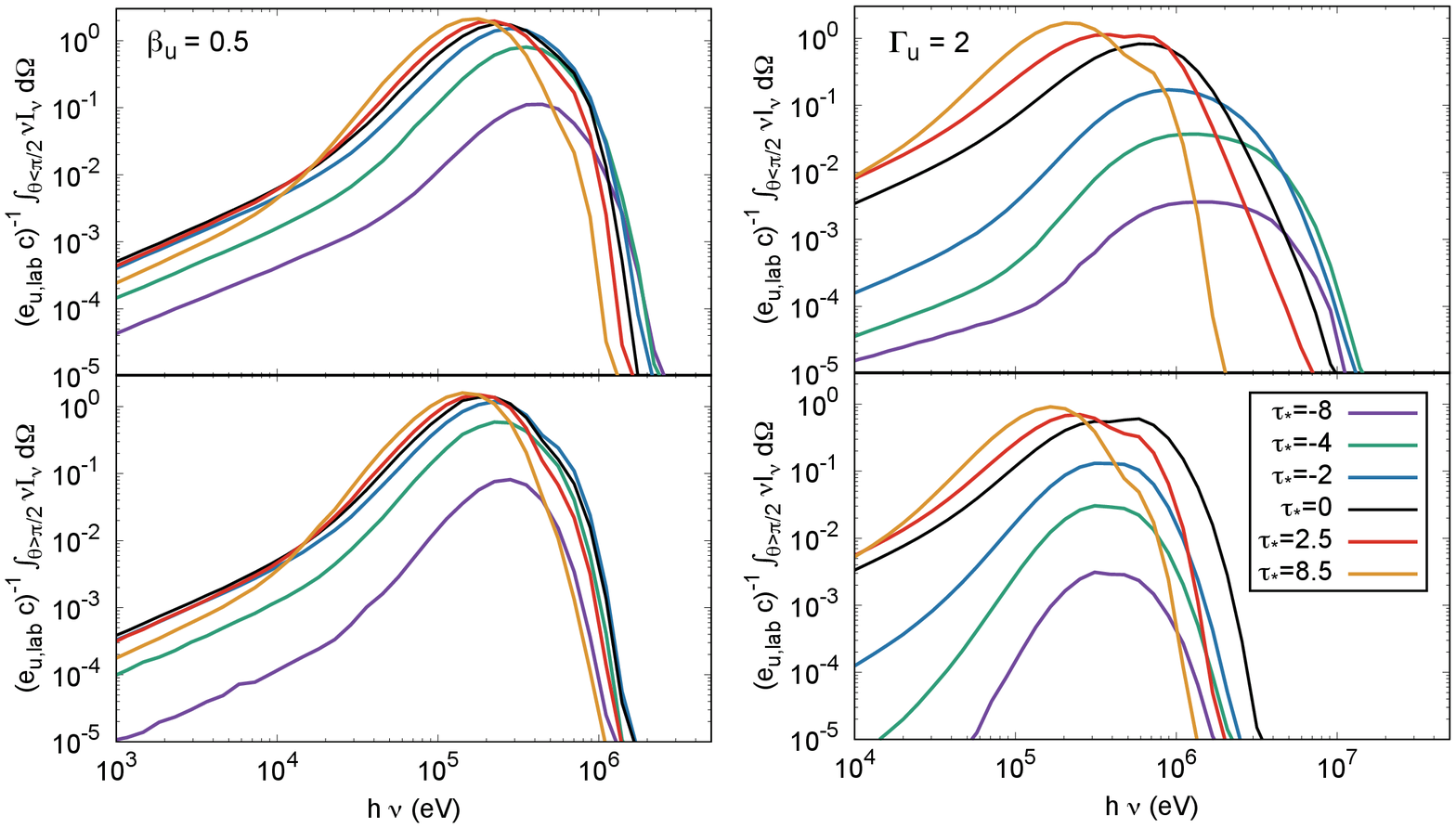}
\end{center}
\caption{Shock frame, local  SEDs, normalized as in Fig \ref{SPDS},
  for $\beta_{\rm u} = 0.5$ ({\it left}) and $\Gamma_{\rm u} = 2$ ({\it right}).
  The top and bottom panels show the photons propagating in an angle contained within a half hemisphere along  ($\theta \leq \pi / 2$) and against  ($\theta > \pi / 2$) the flow, respectively.
  Different lines correspond to the different locations at which the SEDs are evaluated, as indicated.
  The magenta, green, blue, black, red and yellow correponds to $\tau_* = -8$, $-4$, $-2$, $0$, $2.5$ and $8.5$, respectively.
}  
\label{InuB5G2}
\end{figure*}

Fig. \ref{SPDS} displays the angle-integrated spectral energy distribution (SED), as seen in the shock frame, at two locations in the immediate downstream, as indicated.
(The angle dependent SED is shown for illustration in Fig. \ref{SPDS_diff} for $\Gamma_{\rm u}=20$ at $\tau_*=2.5$.)
All spectra exhibit substantial deviations from black body, as expected for fast RMS having $\beta_{\rm u}>0.05$ (see discussion above).   The portion of the spectrum below the peak is 
much softer than that of a Planckian ($\nu I_\nu \propto \nu^3$) in all cases.  
It is produced by thermal Comptonization of soft photons that are continuously generated by bremsstrahlung emissions.
The transition to the Planck regime occurs at a frequency (seen here only for $\beta_{\rm u}=0.1$) 
below which absorption becomes fast enough.   This break frequency generally increases with decreasing downstream temperature (or shock velocity), and
for the spectra exhibited in Fig. \ref{SPDS} is about $20$ eV for $\beta_{\rm u}=0.1$ and  $1$ eV for $\beta_{\rm u}=0.5$.
The overall spectrum slowly evolves towards a black body spectrum as the radiation is advected away from the shock, but full thermodynamic equilibrium is established
only relatively far downstream, as demonstrated in Fig. \ref{InuB5G2} for mildly relativistic shocks (see also Fig. \ref{B1SEDcomp} for $\beta_{\rm u}=0.1$).   This can greatly affect the detection rate of  fast Newtonian and 
relativistic breakouts, since the flux in the optical band during the breakout phase is larger by up to several orders of magnitudes than that naively anticipated 
by invoking a Wien spectrum (see Section \ref{application} below).

The spectrum above the peak is well fitted by an exponential cut-off for $\beta_{\rm u} = 0.1$, but exhibits an extension in the relativistic regime (already
noticeable at $\beta_{\rm u}=0.5$, see Fig. \ref{InuB5G2}), that becomes increasingly more prominent at increasingly larger Lorentz factors, extending up to $\sim \Gamma_{\rm u}^2 m_ec^2$.
The origin of this power law tail  is bulk Comptonization of counterstreaming photons in the deceleration zone. 
However, this high energy component is strongly beamed along the flow (see Fig. \ref{SPDS_diff}), and is present only in 
a beam that subtends an angle of $\sim 1/\Gamma_{\rm u}$ around the flow direction.    As a consequence, it is unlikely to be seen in highly relativistic breakouts 
(since we observe the counterstreaming photons that escape through the upstream region).  However, it might have some effect on the observed spectrum
in mildly relativistic breakouts from aspherical shocks.
    It should be also noted that, while the beamed component is difficult to be observed, certain extension from exponential cut-off is likely to be observed
    even in the spherical breakout for $\beta_{\rm u} \gtrsim 0.5$.
    This can be confirmed in the lower panels of  Fig. \ref{InuB5G2} which show the spectra of counterstreaming photons.

\subsection{Comparison with previous works}

\begin{figure}
\begin{center}
\includegraphics[width=8.5cm,keepaspectratio]{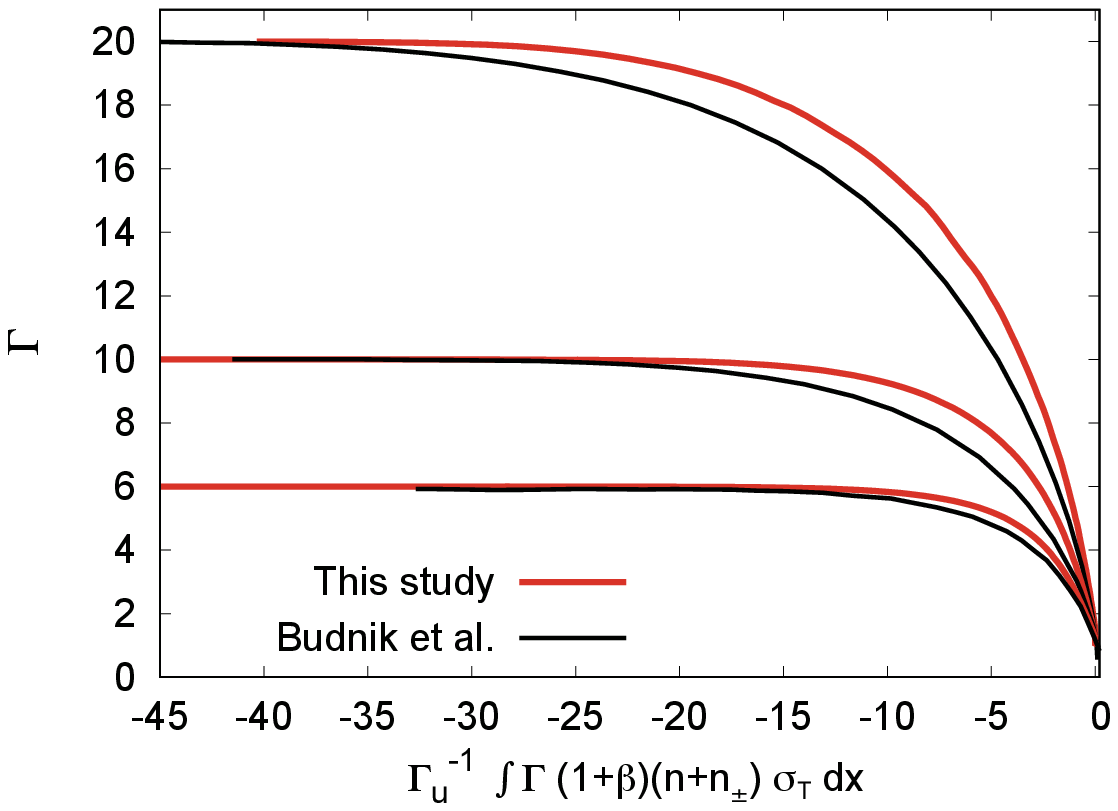}  \includegraphics[width=8.5cm,keepaspectratio]{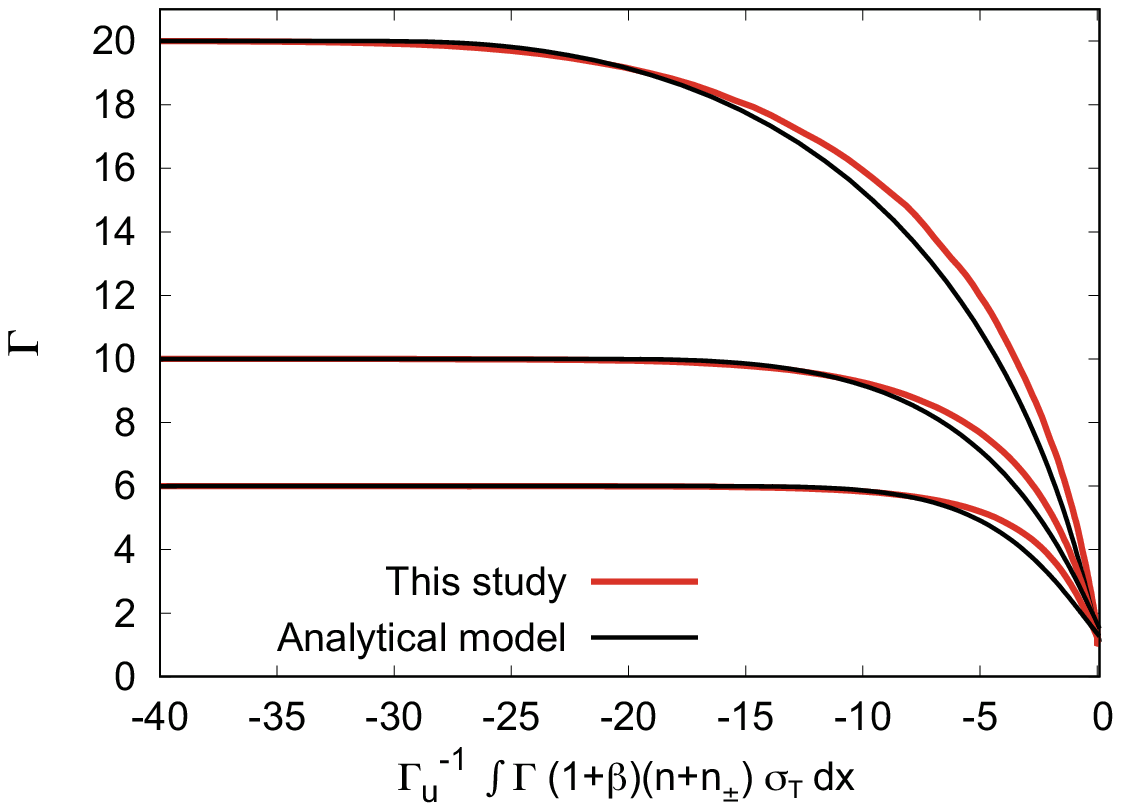}
\end{center}
\caption{Comparison of the Lorentz factor profiles obtained din the simulation 
for $\Gamma_{\rm u} = 6$, $10$ and $20$ with the numerical results of \citet{BKSW10} (upper panel) 
and the analytical solutions of GNL18 (bottom panel).  
The profiles are given here as functions of the dimensionless variable $\Gamma_{\rm u}^{-1} \int \Gamma (1+\beta) (n + n_{\pm}) \sigma_T dx$ used in \citet{BKSW10},  
which different from the pair loaded Thomson depth $\tau_{*}$ used in Figs  \ref{ALL} and \ref{SUB}.} 
\label{G_BUD}
\end{figure}

As a check on our results, we compared the shock structure obtained in the simulations with analytical and numerical solutions reported in the literature.  
In the upper panel of Fig. \ref{G_BUD} we show a comparison of the Lorentz factor profiles obtained in our simulations for  $\Gamma_{\rm u}=6, 10$ and $20$ with those 
computed numerically by \citet{BKSW10}.
As seen, broad agreement is found in all cases.
It should be noted, however, that our simulations systematically find somewhat steeper profile (faster deceleration).
One possible reason for this  discrepancy might be the optimaization of the cross sections in their numerical analysis (see Appendix \ref{DIFFERENCE} for details).
We stress that our code employs the full Klein-Nishina cross sections for Compton scattering and pair production, thus likely producing more accurate results.
Moreover, the current simulations have advantage in that we cover a larger computational domain to avoid any effects related to boundary conditions. 
We also find that the flexibility of the Monte-Carlo method enables us to resolve the momentum distribution of photons with 
a higher precision since we inject sufficiently large number of particles\footnote{In each simulation, more than $10^{9}$ particles are injected.} to minimize  statistical errors.

Next, let us consider the comparison with the analytical solution derived in GNL18.
There are two free parameters in this model  ($\eta$ and $a$) that reflect uncertainties associated 
with the distribution of photons within the shock and the fine details of energy deposition.  
We find that good fits can be obtained both to \cite{BKSW10} results and to our results (see lower panel in Fig. \ref{G_BUD}) for
the range of parameters $\eta = 0.45-0.55$ and $a = 1.5 -2.5$, despite the differences mentioned above.
This suggests that  these differences are not due to a drastic change in the properties of shock.

Analytical solutions have also been derived for sufficiently subrelativistic infinite shocks \citep{BP81b}, as well as
finite shocks with escape \citep{IL19}, by employing the diffusion approximation \citep[e.g.,][]{BP81a}.  
In Appendix \ref{NR} (Fig. \ref{NRcomp}) we compare the shock profile obtained in the simulation for $\beta_{\rm u}=0.1$
with the analytic solution of \cite{BP81b} and find remarkable agreement, confirming the validity of the  diffusion approximation up to at least $\beta_{\rm u}=0.1$.

Finally, the spectra computed in our simulations for shocks  with $\Gamma_{\rm u} \geq 6$ are generally consistent with those obtained by \citet{BKSW10},
although there are some small differences in the spectral evolution just behind the shock.   In particular, the emergence of the thermal bump seen
in Fig. \ref{SPDS}  by comparing the spectra at $\tau_*=0$ and $\tau_*=2.5$, which results from up-scattering of soft photons by the thermal pairs, appears to be
faster in our simulations.   This difference is most likely due to the larger strength of the subshock in our simulations, that gives rise to a higher temperature 
behind the subshock.  At any rate, this difference does not affect the spectrum of the breakout signal (which is emitted in forward direction) in most situations.

\section{Applications to shock breakout emission}
\label{application}
The first signal observed during a shock breakout episode is emitted from a layer of optical thickness $\sim 1/\beta_{\rm d}$ (roughly one diffusion length)
behind the shock \citep{W76,KBW10,NS10,NS12,LN19}.  In sudden breakouts of fast Newtonian RMS  from a sharp density gradient 
(e.g., from a stellar envelope or the fast tail of merger ejecta), the spectrum of the 
breakout emission should closely resemble the spectrum computed in our simulations of infinite RMS.    Our results indicate that the observed temperature
in such breakouts should largely exceeds the black body limit (by up to several orders of magnitudes), and that the spectrum below the peak is
very soft ($F_\nu\propto \nu^0$  roughly), with a prominent bump near the peak (see Figs. \ref{SPDS} and \ref{InuB5G2}).   GRB 060218, discussed below, may be an example.
In relativistic breakouts, in which the opacity is dominated by pairs, the breakout emission is released only after the plasma cools to a temperature
at which the pairs disappear \citep{NS12}.   From our results we estimate  a comoving temperature of about $35$ keV at breakout, and an observed temperature 
of $35\Gamma_{\rm u}$ keV.  Also in this case the spectrum below the peak should be very soft.

In case of shock breakout from a stellar wind the shock structure and spectrum are expected to evolve during the breakout episode due to 
radiative losses (GNL18, \citealt{IL19}), and further analysis is needed to make firm predictions.  Such an analysis is underway. 

In the following we consider some specific systems:

\subsection{GRB 060218}
GRB 060218 is one of the six low luminosity GRBs identified thus far \citep{campana2006,SKN06}.  It is associated with SN 2006aj - a rare type double peak SN.
It has long been proposed \citep{kulkarni1998} that, unlike regular LGRBs,  llGRBs may result from a breakout of a shock driven by a choked jet, and
this model has been applied to explain the properties of GRB 060218 \citep{campana2006,waxman2007,li2007,nakar2015}.   
It has been argued \citep{nakar2015} that the association of LGRBs and llGRBs with 
the same rare SN type (broad line IC), suggests that both GRB classes share the same explosion physics,  but in different environments.   
Specifically, \cite{nakar2015} proposed that the key difference between llGRBs and LGRBs is the outer structure of the progenitor;
while in llGRBs the compact core of the progenitor is ensheathed by an extended ($> 100 R_{\odot}$), low mass ($\sim 10^{-2} M_\odot$) envelope 
that chokes the jet, in LGRBs this envelope is absent.   The double-peak light curve of SN 2006aj is naturally explained in this model;
the first peak is associated with the cooling emission of the low mass
extended envelope and the second peak is powered by radioactive decay of $^{56}$Ni.  
Furthermore, the envelope parameters needed to account for the optical/UV emission of SN 2006aj around the first peak 
are in remarkable agreement with those needed, independently, to explain the breakout gamma-ray emission.

The shock breakout scenario for GRB 060218 has been criticized  recently  \citep{Em19} on the grounds that it cannot account for the 
UV/optical spectra observed in the first $1350$ s (roughly the breakout duration in the shock breakout model), that appear to be much softer than 
a Rayleigh-Jeans power-law (consistent with flat spectrum, $F_\nu\propto \nu^0$, within the errors).   Moreover, by the time of $\sim 2000$ s
the spectrum evolved into a Rayleigh-Jeans spectrum.    \cite{Em19} therefore proposed that the early ($< 1300$ s) UV/optical emission is produced
by synchrotron emission in an external shock driven by a successful, low power jet, that also produce the GRB emission.  However, this appears inconsistent
with the requirement that the jet penetrates all the way through the extended envelope that is needed to account for the early 
cooling emission of SN 2006aj, unless an engine life time in excess of 10,000 s is invoked. 

Here we argue that the observed UV/optical spectrum at $t<1000$ s is in fact consistent with the shock breakout model. 
Our results indicate that the observed peak energy of the GRB ($\sim$ 40 keV, \citealt{KRG07}) can be accommodated by a shock velocity of $\beta_{\rm u}\sim 0.3-0.4$
(depending somewhat on shock
 geometry and other details).  This velocity, in turn, yields breakout energy and duration that are consistent with the observed 
 GRB isotropic equivalent energy and duration  \citep{nakar2015}.   From our simulations we find that for a shock velocity in this range the portion of the spectrum 
below the peak is roughly flat, down to a few eV, consistent with the UV/optical slope reported by \cite{Em19}.   The luminosity in the V band
is lower by a factor of $\sim 10^4$ than the GRB luminosity at $650$ s, consistent with a roughly flat spectrum up to the X-ray band,
as predicted by the RMS simulations.  The evolution to the Rayleigh-Jeans regime at $t >1600$ s is also naturally expected, since 
the emission at these times originate form deep layers behind the shock, in which the black body limit has established.

\subsection{GRB170817A}
The shock breakout model for NS mergers \citep{kasliwal2017,gottlieb2018,pozanenko2018,beloborodov2018} asserts that the gamma-ray flash observed in GRB 170817A was produced during 
the emergence of the shock driven by the jet-cocoon system from the fast tail leading the merger ejecta.   The original model proposed by \cite{kasliwal2017}
and \cite{gottlieb2018}
has been criticized by \cite{BLL18} on the grounds that it predicts a too high SED peak for the ejecta Lorentz factor needed to accommodate observational
constraints.  Here we point out that, contrary to this claim, the  sensitive dependence of the temperature on shock velocity allows 
sufficient freedom to consistently account for all observables.  As shown recently  \citep{N19}, the model can 
reproduce the burst energy and duration, given the observed delay between the NS collision and the onset of the burst, for shock
velocities in the range $0.25 \lesssim \beta_{\rm u}  \lesssim 0.6$.  According to our simulations, this corresponds to the range of comoving temperature
at breakout of 10  keV $\lesssim kT \lesssim 35$ keV (the upper limit is the temperature at which pairs disappear).   
With an ejecta Lorentz factor of $\Gamma=5$ invoked in \cite{BLL18} this gives 
a peak energy in the range $200 - 500$ keV, consistent with the observations.  We stress that our estimate assumes H-rich composition.  
Heavy composition, particularly r-process, would imply a lower temperature, but still in a range consistent with the observations \citep{N19}.

\section{Summary and conclusions}
\label{conclusion}

We performed Monte-Carlo simulations of photon-starved RMS  over a broad range of shock 4-velocities.
In these shocks, the (cold) upstream flow decelerates via bulk Comptonization of counterstreaming
photons that are generated inside and just behind the shock by bremsstrahlung emission.
Six models, that cover the  fast Newtonian ($\beta_{\rm u} = 0.1$), trans-relativistic ($\beta_{\rm u} = 0.5$), mildly relativistic ($\Gamma_{\rm u} = 2$) and
fully relativistic ($\Gamma_{\rm u} = 6, 10$ and $20$) regimes are thoroughly investigated.
All models invoke a pure H composition at a fiducial density of $n_{\rm u}=10^{15}$ cm$^{-3}$ in the far upstream flow.   
This is the first time that self-consistent calculations of the shock structure and spectrum in the fast Newtonian and mildly relativistic regimes ($\Gamma_{\rm u}<6$) are performed.
Our results for fully relativistic  ($\Gamma_{\rm u} \ge 6$) RMS are in good agreement with the numerical solutions obtained by \cite{BKSW10}, and the analytical 
solutions derived by \cite{NS12} and \cite{GNL18} for the shock profile.    The main findings are:

(i) Our simulations confirm that in the fast Newtonian regime ($0.05 \lesssim \beta_{\rm u} \lesssim 0.5$) the immediate downstream 
temperature depends sensitively on shock velocity (roughly as $\beta_{\rm u}^3$), whereas in relativistic shocks it is regulated by exponential pair creation, and lies 
in the range $100 - 200$ keV, with a very weak dependence on $\Gamma_{\rm u}$.  For the assumed density and composition (pure H), the transition to pair 
dominance was found to occur at $\beta_{\rm u}=0.5$, as anticipated earlier.  For r-process composition it is expected to occur at somewhat higher velocity
 \citep{LN19}.   In Section \ref{application} we discussed the implications of the sensitive dependence of $T$ on $\beta_{\rm u}$ for the shock breakout model of GRB 170817A.

(ii) In all cases explored ($\beta_{\rm u} \ge 0.1$) the radiation in the immediate downstream is out of thermodynamic equilibrium due to insufficient photon generation.
The black body limit is reached only relatively far downstream.   As a result, the spectrum below the $\nu F_\nu$ peak is considerably softer than the Planck spectrum
down to a break frequency that depends on shock velocity, below which the spectrum hardens ($F_\nu\propto \nu^2$).   This implies a much brighter optical emission 
in fast Newtonian and relativistic breakouts than naively estimated by invoking Wien spectrum below the peak, with important consequences for the detection rate of 
shock breakout candidates.   In particular, we argued that the softening of the spectrum below the peak is consistent with the 
early X/UV/optical emission detected in GRB 060218.   A detailed analysis of the observational consequences is underway.

(iii)  In relativistic shocks the photon distribution inside the shock is highly anisotropic.  For the photon beam that subtends 
an angle $\sim 1/\Gamma_{\rm u}$ around the flow direction (that is, moving towards the downstream), 
the spectrum above the peak extends to an energy of $\Gamma_{\rm u}^2 m_ec^2$ in the shock frame.  This should not affect the observed spectrum in 
most relativistic breakouts, but might have some effect on the high energy spectrum in mildly relativistic, aspherical breakouts,
which are expected in cases where the shock is driven by a jet as, e.g., in BNS mergers and llGRB.  The reason is that an observer
located at some angle to the axis will receive contributions from different parts of the shock, each moving at a different Lorentz factor and
in a different direction.

(iv) While in fast and mildly relativistic RMS the shock transition is smooth, relativistic RMS ($\Gamma_{\rm u} \ge 2$) exhibit subshocks 
with a local velocity jump of  
$\delta(\Gamma\beta)\sim0.2 - 0.7$
for $\Gamma_{\rm u} = 2 - 20$,
and a non-negligible strength.  We find that 
a few percents of the total shock energy dissipate in the subshock.  It is unclear at present wether these subshocks 
can accelerate particles to highly relativistic energies, but if they can it might significantly affect the emitted spectrum.   Further investigation is needed to quantify
the effect of the subshock on the high energy emission.

Our simulations provide an important insight into the properties of fast and relativistic RMS, and their role in shaping the 
shock breakout signal in energetic supernovae, low luminosity GRBs and NS mergers.   The results of our simulations 
can be employed to predict the spectral evolution during the breakout phase under conditions anticipated in specific systems. 
However, the present analysis applies to infinite shocks and may not be adequate enough to describe breakouts from stellar winds,
in which radiative losses become gradually important during the breakout phase, changing  the shock structure  (GNL18, \citealt{IL19}).
Our Monte-Carlo technique has been generalized recently to finite RMS with radiative losses, and the investigation of such shocks 
is currently in progress, and will be reported  in a future publication (Ito, Levinson \& Nakar, in preparation). 

 \section*{Acknowledgments}
 We thank Ehud Nakar for enlightening discussions.  This work was supported by JSPS KAKENHI Grant Number JP16K21630, JP16KK0109, JP19K03878 and JP19H00693.
 Numerical computations and data analysis were carried out on XC50  at Center for Computational Astrophysics, National Astronomical Observatory of Japan, Hokusai BigWaterfall system at RIKEN and the Yukawa Institute Computer Facility.
This work was supported in part by a RIKEN Interdisciplinary Theoretical \&
Mathematical Science Program (iTHEMS) and a RIKEN pioneering project ``Evolution of Matter in the Universe (r-EMU)'' and
``Extreme precisions to Explore fundamental physics with Exotic particles (E3-Project)''.  AL acknowledges support by the Israel Science Foundation Grant 1114/17. 
 

\appendix
\section{IMPLEMENTING BREMSSTRAHLUNG EMISSION AND ABSORPTION}
\label{sec:appA}

The Monte-Carlo  code used in this study
handles transfer of photons in a medium at which Compton scattering, 
pair creation/annihilation, and bremsstrahlung emission/absorption take place.
The description of the original code that includes the former two processes is given in Paper I.
Here we outline the  implementation of the bremsstrahlung process in the modified code.
Following the notations in Paper I, we label quantities that are measured in the comoving frame
of the bulk plasma with the superscript prime symbol.

\subsection{Bremsstrahlung emission}
The modified code includes photon production by $e^\pm p$, $e^-e^+$, $e^-  e^-$ and $e^+ e^+$ thermal bremsstrahlung.
The photon generation rate in the fluid rest frame is given by \citep{S82}
\begin{eqnarray}
\label{bremss}
\left( \frac{dN_{\gamma}}{dt^{'} d\nu^{'} d\Omega^{'} dV^{'}} \right)_{\rm ff} = \sqrt{\frac{1}{6\pi^3}} ~\alpha_{f} \sigma_T c n^2 \frac{1}{\nu^{'} \sqrt{\Theta}}~ {\rm exp}\left( - \frac{h \nu^{'}}{k T}\right) \lambda_{\rm ff}, 
\end{eqnarray}
where $\Theta = k T / m_e c^2$  denotes the plasma temperature in units of the electron mass, and $\alpha_f$ is the fine structure constant.\footnote{The different pre-factor 
that appear in equation (57) of \citet{BKSW10} is due to the difference in the definition of $\lambda_{\rm ff}$. While we directly use the Gaunt factors given in \citet{SDR95}, \citet{BKSW10} multiplies them by a factor $\pi / 2\sqrt{3}$.}
Here $n$ denotes the baryon density, and
\begin{eqnarray}
\lambda_{\rm ff}  = (1+ 2 x_{+})g_{ep} + [x_+^2 + (1 + x_+)^2] g_{ee} + x_+ (1+ x_+) g_{\pm}
\end{eqnarray}
is a dimensionless factor that depends on the pair-to-baryon density ratio, $x_{+} = n_{\pm}/2n$,
and the Gaunt factors $g_{ep}$, $g_{ee}$ and $g_{\pm}$ that correspond to $e^\pm p$, $e^\pm e^\pm$ and $e^- e^+$ encounters, respectively.
As for the Gaunt factors, we employ the analytic fits given in \citet{SDR95}
which are expressed as functions of temperature and frequency. Note that there is an errata for this paper \citep{SDR96}.

The photon generation rate in the shock frame is related to the rate in the fluid frame through
\begin{eqnarray}
\label{VV}
\left( \frac{dN_{\gamma}}{dt d\nu d\Omega dV} \right)_{\rm ff}   =
     {\cal D} \left( \frac{dN_{\gamma}}{dt^{'} d\nu^{'} d\Omega^{'} dV^{'}} \right)_{\rm ff},
\end{eqnarray}     
where ${\cal D} = [\Gamma (1 - \beta  {\rm cos}\theta)]$ is the Doppler factor.
In our code,  photon packets are injected at every grid point in the numerical domain  with a probability proportional to the local photon generation rate.
    We employ a fixed value for the photon number per packet, $n_{\gamma, {\rm pack}}$\footnote{The photon number per packet is also fixed for the photons produced by pair annihilation, but the value is defferent from that employed for the bremsstrahlung emission.
        It is also noted that single packet is splitted into multiple packets that contain smaller number of photons after scattering when appreciable increase in energy takes place.
      This is done to avoid low photon statistics at high energy.},
    in all grid points, and  
    the injecting rate of the packet number per unit volume
for a given range of solid angles  $d\Omega$ and frequencies $d\nu$ is given by
$\int \int \left( \frac{dN_{\gamma}}{dt d\nu d\Omega dV} \right)_{\rm ff} (n_{\gamma, {pack}})^{-1}d\Omega d\nu$.

Since the photon generation rate scales as $\left( \frac{dN_{\gamma}}{dt d\nu d\Omega dV} \right)_{\rm ff} \propto \nu^{-1}$
in the low frequency limit ($\nu \ll kT$), the produced number of photons diverges logarithmically at low energy.\footnote{In reality, the photon number at a given frequency will be limited by that of black body due to the absorption process.}
Therefore, we must impose a lower limit on the frequency that we take into accout in order to avoid divergence.
Here, we follow the prescription employed in \citet{BKSW10} for the minimum frequency which is determined from 
\begin{eqnarray}
\nu_{min} = \frac{\gamma_{e, th}^2 \beta_{e, th} c}{2 \pi \lambda_D}, 
\end{eqnarray}
where $\lambda_D = \sqrt{kT / 4 \pi e^2 (n + 0.5 n_{\pm})}$ is the Debye length
and $\gamma_{e, th} = 1 + 3/2 f(T) \Theta$ and $\beta_{e,th}= \sqrt{1-\gamma_{e, th}^{-2}}$ are the Lorentz factor and velocity of the theral motion of electrons, respectively.
Here $f(T) = {\rm tanh}[({\rm ln}\Theta + 0.3) / 1.93] + 3/2$ is an analytical function of temperature defined in \citet{BKSW10}, obtained from a fit to the exact  equation of state of pairs.
Photons at low frequency is produced at impact factor larger than the Debye length, and therefore the bremsstrahlung emission is suppressed due to Coulomb screening.\footnote{The value of low frequency cut-off is important, since the photon generation rate of free-free emission do not converge at low frequencies. It is noted however, that  the effective minimum frequency of the photons that contributes to the increase in the photon number is much higher and is determined by the condition for the photons to be boosted up to thermal peak energy before being re-absorbed \citep[see e.g.,][]{KBW10, LN19}.
Indeed large number of low energy photon packets are quickly absorbed 
      after being injected by free-free absorption.
}

\subsection{Bremsstrahlung absorption}

The opacity for the bremsstrahlung process is determined from the Kirchhoff's law for radiation.
In terms of the photon generation rate it is give by
\begin{eqnarray}
\alpha^{'}(\nu^{'},T) =  \left( \frac{dN_{\gamma}}{dt^{'} d\nu^{'} d\Omega^{'} dV^{'}} \right)_{\rm ff} \frac{h \nu^{'}}{B_\nu (\nu^{'}, T)} 
\end{eqnarray}
in the fluid rest frame, where $T(x)$ is the local temperature, and $B_\nu (\nu, T) = 2 h \nu^3 / c^2 [{\rm exp}(h\nu/kT) - 1]^{-1}$ is the corresponding Planck function.
With the known opacity, the distance a photon packet  propagates before being absorbed is determined in the same manner as in the case of 
Compton scattering and photon-photon pair creation (see paper I). In brief, 
we randomly draw a value for the optical depth, say $\delta \tau$, and then determine the physical distance $l$ in the laboratory frame, using the implicit equation
$\delta\tau = \int^l_0 {\cal D} \alpha^{'} dx$.

\section{BOUNDARY CONDITIONS}
\label{sec:app_boundary}

As for the boundary conditions at far upstream, we employ the same method used in Paper I. We inject photons with photon-to-baryon number ratio given by
$\tilde{n}_{\rm u}=n_{\gamma {\rm u}}/n_{\rm u} = 10$ at the boundary. The energy contained in the injected photons is negligible compared to the baryons ($\tilde{n}_{\rm u} k T_{\rm u}/ m_p c^2 \ll 1$).
The injected spectrum is Wien,  characterized by the  local temperature in the comoving frame.  The temperature at the boundary is set 
    as $T_{\rm u} = (3 k n_{\gamma {\rm u}} / a_{rad})^{1/3}$ under the assumption it is close to the blackbody limit, where $a_{rad}$ is the radiation density constant.
    Note that, while the assumption of blackbody contradicts with the assumption of Wien spectrum,  the spectral feature is only noticable at far upstream region and   is quickly smeared out by the emission/absoption process.
It is emphasized that the current boundary condition is adopted for numerical convenience, and 
has no noticeable effect on the results, since the injected (advected) photon population is highly sub-dominant
compared to the photons that are generated inside the shock.\footnote{Our code is capable of handling much smaller number of photon to baryon ratio at the boundary. However, smaller number of photons implies lower temperature at the boundary which in turn results in a larger temperature increase at far upstream regions. 
    Since capturing such feature slightly enlarges the computational time, the current value 
    is employed for  $\tilde{n}_{\rm u}$.
}

We also inject photons at the far downstream boundary in order to maintain the radiation field there isotropic in the comoving frame, 
with a Wien spectrum which is determined by the local plasma temperature. The normalization of the photon's energy distribution 
is iteratively adjusted  to match the outgoing photos  flux at the boundary.
It is noted that, while the assumption that the radiation is isotropic can be justified, 
the choice of a Wien spectrum at the boundary is approximate, since in practice the spectrum tends to approach the Planck distribution. 
Nonetheless, this should not affect the results  drastically since the downstream boundary is located at a distance from the shock 
which is larger than the diffusion length, \footnote{In all models, the total optical depth of the  downstream regions 
at least factor of $1.5$ larger than the diffusion length ($\tau_{*} > 1.5 \beta_{\rm d}^{-1}$).}
so that photons injected at the downstream boundary cannot reach the shock and, therefore, cannot affect its dynamics.

To sum up, as in Paper I,  the photon flux
 at the upstream and downstream boundaries in the laboratory frame is a function of the
 photon number density and temperature, and can be written as
\begin{eqnarray}
 \frac{dN_{\gamma}}{dt d\nu d\Omega dS} = {\cal D}^{2} \frac{dN_{\gamma}}{dt^{'} d\nu^{'} d\Omega^{'} dS^{'}} , 
\end{eqnarray}
where 
\begin{eqnarray}
 \frac{dN_{\gamma}}{dt^{'} d\nu^{'} d\Omega^{'} dS^{'}} = \frac{n_{\gamma}}{8 \pi}
                               \left(\frac{h}{k T} \right)^3 \nu^{'2} {\rm exp}\left( - \frac{h \nu^{'}}{k T}\right) . 
\end{eqnarray}
Thus, for a given range of solid angles  $d\Omega$ and frequencies $d\nu$, 
 $\frac{dN_{\gamma}}{dt d\nu d\Omega dS} (n_{\gamma, {pack}})^{-1} {\rm cos} \theta d\Omega d\nu$ gives the injection rate of  the packet number per unit area of the boundary surface, where $n_{\gamma, {\rm pack}}$ is the number of photons contained in a single packet.
However, it is again emphasized that the boundary conditions do not effect the solutions,  which
 is governed by the photons that are generated inside the shock and within one diffusion length behind it.

\section{COMPARISON WITH ANALYTIC SOLUTION IN NON-RELATIVISTIC REGIME}
\label{NR}
\begin{figure}
\begin{center}
\includegraphics[width=8cm,keepaspectratio]{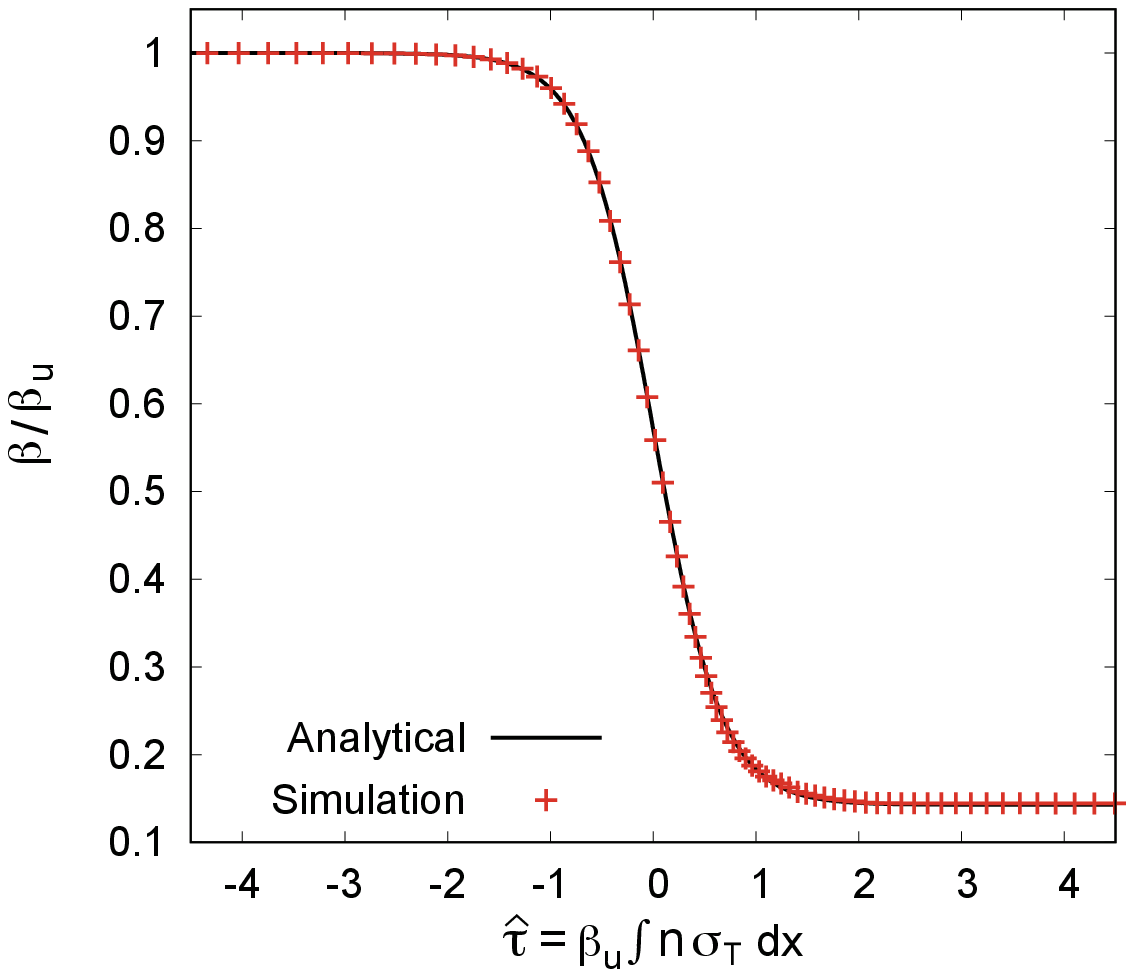}
\end{center}
\caption{Comparison of the velocity structure obtained in our simulation for $\beta_{\rm u} = 0.1$ ({\it red crosses}) and the analytical solution Eq. (\ref{betadiff})
  ({\it black solid line}).
 }
\label{NRcomp}
\end{figure}

Here we compare our simulation for $\beta_{\rm u} = 0.1$ with the analytic solution of a diffusive shock.
When the speed of the upstream flow is well below the speed of light, $\beta_{\rm u} \ll 1$, diffusion approximation 
can be applied to solve the RMS structure.
This approximation yields an universal velocity profile which depends only on an
optical depth defined as ${d\hat{\tau}} = \beta_{\rm u} \int n \sigma_T dx$ \citep[see e.g.,][]{BP81b, KBW10, IL19}:
\begin{eqnarray}
  \frac{\beta}{\beta_{\rm u}} = \frac{4}{7} + \frac{3}{7} {\rm tanh}\left[- \frac{3}{2} \hat{\tau} \right].
  \label{betadiff}
\end{eqnarray}
  Here $\hat{\tau}=0$ is located roughly at the center of the shock $\beta / \beta_{\rm u} = 4/7$, 
  and $\hat{\tau} \rightarrow - \infty$ ($\hat{\tau} \rightarrow + \infty$) and correspond to the far upstream (far downstream).
  The red crosses in Fig. \ref{NRcomp} delineate the result of our simulation and the black solid line the analytic solution.
  As is seen the agreement is remarkable.
  The fact that our code is capable of reproducing the universal profile to such an accuracy 
  reassures that our calculations are highly reliable.

\section{SED EVOLUTION FOR $\beta_{\rm u} = 0.1$}
\label{B1SED}
Here we examine the evolution of the SED at the downstream region for $\beta_{\rm u} = 0.1$.
We remark that while the velocity structure shows universal profile as described in Appendix \ref{NR}, 
The temperature and, consequently, the spectrum depend on the composition and density $n_{\rm u}$ far upstream.
In this section, we consider a pure H composition at a density $n_{\rm u} = 10^{15}~{\rm cm}^{-3}$.

\begin{figure}
\begin{center}
\includegraphics[width=8cm,keepaspectratio]{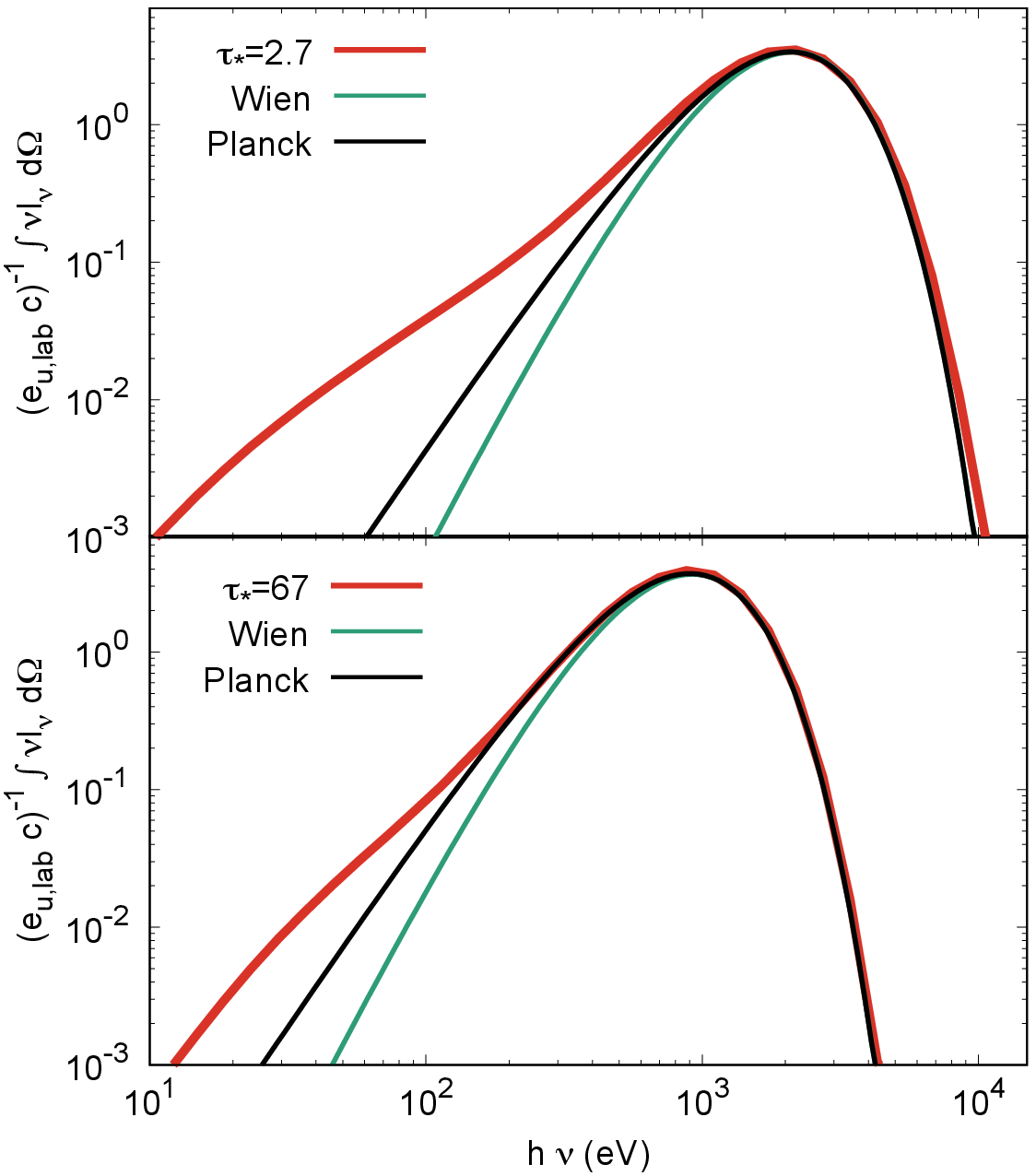}
\end{center}
\caption{Comparison of the local SEDs in the downstream region
  normalized by the energy density of the far upstream $e_{\rm u} = \Gamma_{\rm u} (\Gamma_{\rm u} - 1) n_{\rm u} m_p c^2$ for the model with velocity of $\beta_{\rm u} = 0.1$ ({\it red}) with the Planck ({\it black}) and Wien distribution ({\it green}).  Note that the anisotropy of photons is negligible in this case. 
  The top and bottom panels correspond to the locations $\tau_{*} = 2.7$ and $\tau_{*} = 67$, respectively.
In the top (bottom) panel, the temperature of the Planck and Wien distributions is $k T = 530~{\rm eV}$ ($k T = 230~{\rm eV}$).
}
\label{B1SEDcomp}
\end{figure}

As stated in the main text, under these conditions  the radiation in the immediate downstream, in the $\beta_{\rm u} = 0.1$case, 
is marginally out of thermodynamic equilibrium.  Our simulation yields a plasma temperature of $k T \sim 520~{\rm eV}$ at
the end of the deceleration zone ($\tau_{*} = 0$),
which is factor of a few higher than the black body limit ($k T_{\rm d} \simeq 130~{\rm eV}$).
At the downstream region, thermalization proceeds owing to continuous photon generation by the bremsstrahlung emission.
As a result, the plasma cools and the spectrum evolves towards a Planckian. 
For illustration, we plot  in Fig. \ref{B1SEDcomp} the spectrum (red line) at two different downstream locations, just behind the shock  ($\tau_*=2.5$) and at one diffusion length away ($\tau_*=67$).\footnote{Note that since $\beta_{\rm d}=\beta_{\rm u}/7$, the optical thickness over one diffusion length downstream is $\tau_*\approx 1/\beta_{\rm d}=70$.}
A slow evolution into a Planck spectrum (shown as the black line) is evident, and it is clear that a full thermodynamic equilibrium will be established
only at a few diffusion lengths downstream.   The important implication is that the spectrum of the breakout emission should be much softer than Planck. 

\section{DIFFERENCE FROM \citet{BKSW10}}
\label{DIFFERENCE}

In this appendix we compare some technical aspects between our simulation method and that of \cite{BKSW10}, in an
attempt to elucidate the origin of the differences between the results of the two simulations
indicated in the main text.

The two  codes incorporate exactly the same radiation interactions (Compton scattering, pair production/annihilation and bremsstrahlung emission/absorption).
However, while our method use the exact cross sections for all reactions, \cite{BKSW10} use some approximations to compute Compton scattering.
First, they assume that the scattering is isotropic in the fluid rest frame, which is inaccurate in the Klein-Nishima regime.
Second, they use an approximate form for the comoving energy redistribution function of scattered photons,  that keeps a relativistic Wien distribution invariant under scattering. 
As shown below, this form is not precise since it is not based on the exact cross section.   The cumulative effect of their approximations is an overestimate of
the mean energy deposition inside and upstream of the shock.   While we cannot completely rule out other causes, we tend to believe that this overestimate is the reason for
the quantitative difference in temperature and velocity profiles. 

\begin{figure}
\begin{center}
\includegraphics[width=8cm,keepaspectratio]{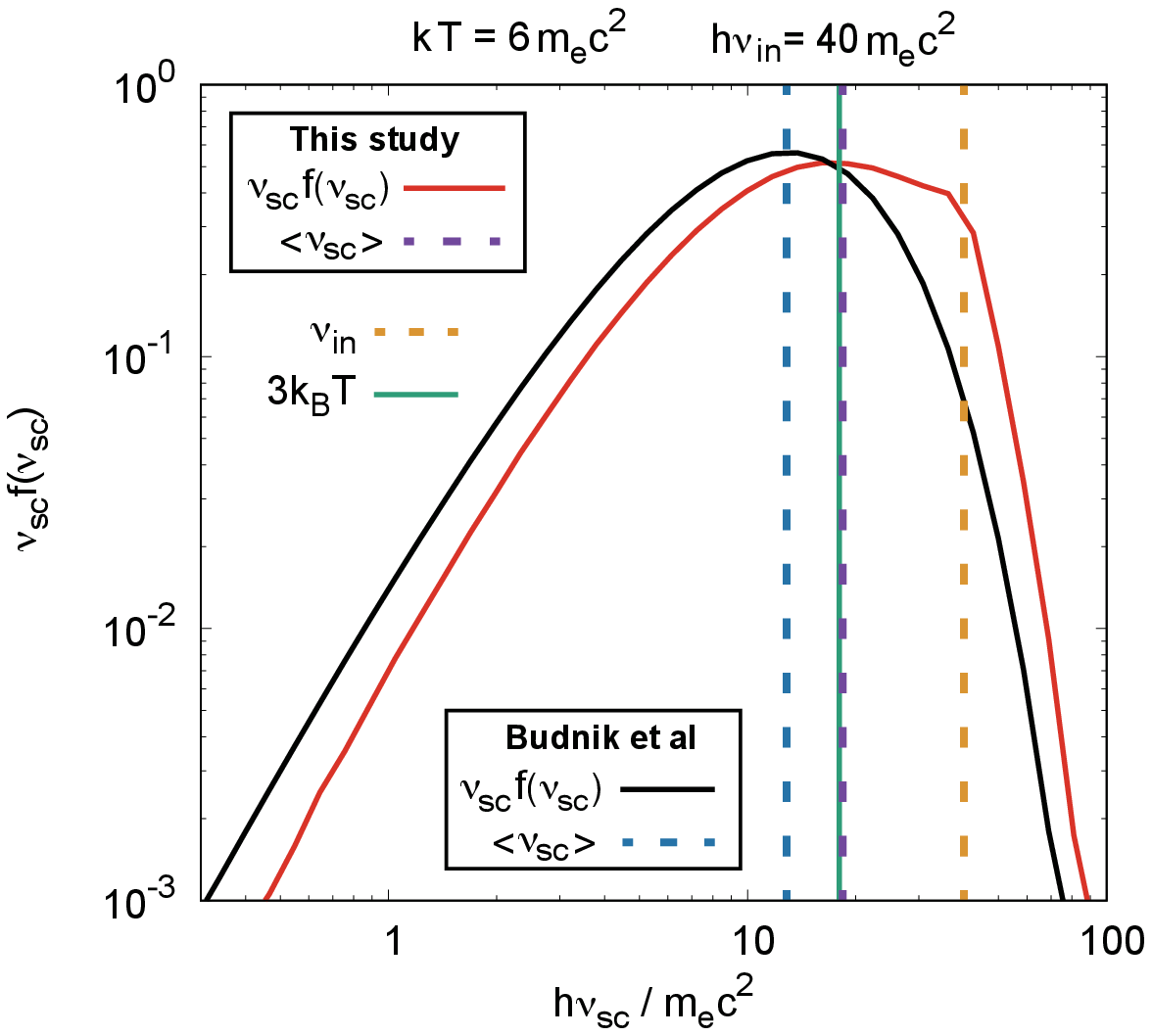}  
\end{center}
\caption{Energy redistribution function of scattered photons for incident photon energy of $h\nu_{\rm in} = 40m_ec^2$, injected in a thermal pool of electrons having a temperature $k T = 6m_ec^2$. Here the redistribution function $f(\nu_{\rm sc}) d\nu_{\rm sc}$ gives the probability for the scattered photon energy to be in the range $[\nu_{\rm sc}: \nu_{\rm sc} + d\nu_{\rm sc}]$.
  The red solid line shows the exact function obtained from our calculations, while the black line shows the analytic function employed in \citet{BKSW10}.
  The dashed lines represent the incident photon energy ({\it brown}), and the average scattered photon energy obtained in our simulations ({\it magenta})  
  and in \citet{BKSW10} ({\it blue}).
 The green solid line marks the typical energy of electrons, $3 kT$, and is shown for reference.
}
\label{RECOIL}
\end{figure}

A comparison of the redistribution function adopted in \cite{BKSW10} and the exact one computed in our simulation by using the full cross section is exhibited 
in Fig. \ref{RECOIL}.
Here the incident comoving photon energy and the plasma
temperature are taken to be $h \nu_{\rm in} = 40 m_e c^2$ and $kT = 6 m_e c^2$, respectively,
which are typical values for the backstreaming photons and the plasma temperature at the deceleration zone in the $\Gamma_{\rm u} = 20$ simulation;\footnote{
We find that the peak energy of the SED of backstreaming photons is $\sim 2 m_e c^2$ with respect to the shock frame. 
In the fluid frame it is boosted to $\sim 40(\Gamma_{\rm u}/20) m_e c^2$.}   the difference is clear.  In particular, the mean energy of scattered photons is
larger in our simulations.  This discrepancy is qualitatively similar for other incident energies and plasma temperatures. 
Consequently, the calculations of \citet{BKSW10} tend to
systematically underestimate the population of high energy photons, implying that the energy gain of the plasma (heating) per 
scattering is overestimated in their simulation.

\begin{figure}
\begin{center}
\includegraphics[width=8cm,keepaspectratio]{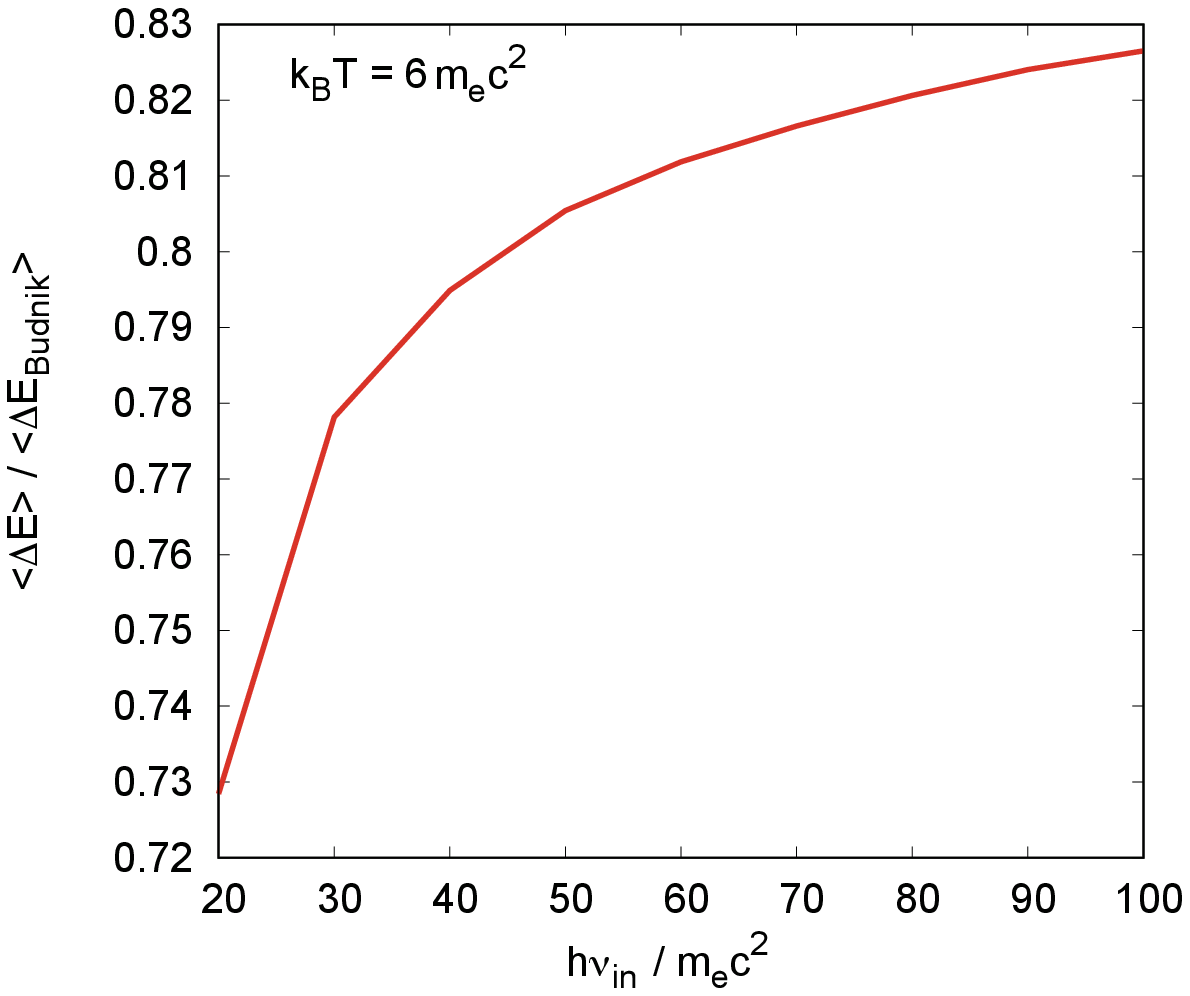}  
\end{center}
\caption{Ratio between the exact value of the mean energy deposition per scattering obtained in our simulations, $<\Delta E >$,
and the value computed in \citet{BKSW10}, $<\Delta E_{\rm Budnik}>$.  The average $<\Delta E>$ is taken over the distribution
of incident photons $f(\nu_{\rm in}$), each deposit a mean amount  $\Delta E = h(\nu_{\rm in} - <\nu_{\rm sc}>)$ in a single scattering.
}
\label{RECOIL_DEP}
\end{figure}

In Fig. \ref{RECOIL_DEP} we plot the ratio between the values of the average energy deposition per scattering obtained  in our simulation and in those of \cite{BKSW10}.
As seen, for $h\nu_{\rm in} > 3 kT$ the average energy deposition per scattering computed in \cite{BKSW10}  is larger by $\sim 20 - 30{\%}$.
While these systematic errors are modest, they accumulate in a nonlinear manner that can somewhat  alter the shock profile. 

\label{lastpage}

\end{document}